\documentclass{elsart}
\usepackage[latin1]{inputenc}
\usepackage[T1]{fontenc}
\usepackage{natbib}
\bibpunct{[}{]}{;}{a}{,}{,}
\usepackage[normalem]{ulem}
\usepackage{verbatim}
\usepackage{graphicx}
\usepackage{amssymb}

\journal{European Journal of Mechanics B/Fluids }
\begin{document}
\begin{frontmatter}
\title{Simulation of an optically induced asymmetric deformation of a liquid-liquid interface}

\author[TREFLE]{Hamza Chra\"ibi  \corauthref{cor}},
\corauth[cor]{Corresponding author.} \ead{h.chraibi@cpmoh.u-bordeaux1.fr}
\author[TREFLE]{Didier Lasseux},
\author[TREFLE]{Eric Arquis},
\author[CPMOH]{R\'egis Wunenburger},
\author[CPMOH]{Jean-Pierre Delville},

\address[TREFLE]{Universit\'e Bordeaux I\\Transferts, \'Ecoulements, Fluides, \'Energ\'etique (TREFLE), UMR CNRS 8508\\
Esplanade des Arts et M\'etiers\\
33405 Talence Cedex, France.}

\address[CPMOH]{Universit\'e Bordeaux I\\Centre de Physique Mol\'eculaire Optique et Hertzienne (CPMOH), UMR CNRS 5798\\
351 Cours de la Lib\'eration\\ 33405 Talence cedex, France.}

\begin{abstract}

Deformations of liquid interfaces by the optical radiation
pressure of a focused laser wave were generally expected to
display similar behavior, whatever the direction of propagation of
the incident beam.

Recent experiments showed that the invariance of interface
deformations with respect to the direction of propagation of the
incident wave is broken at high laser intensities. In the case of
a beam propagating from
 the liquid of smaller refractive index to that of larger one, the
interface remains stable, forming a nipple-like shape, while for
the opposite direction of propagation, an instability occurs,
leading to a long needle-like deformation emitting micro-droplets.
While an analytical model successfully predicts the equilibrium
shape of weakly deformed interface, very few work has been
accomplished in the regime of large interface deformations. In
this work, we use the Boundary Integral Element Method (BIEM) to
compute the evolution of the shape of a fluid-fluid interface under the
effect of a continuous laser wave, and we compare our numerical
simulations to experimental data in the regime of large
deformations for both upward and downward beam propagation. We
confirm the invariance breakdown observed experimentally and find
good agreement between predicted and experimental interface hump
heights below the instability threshold.
\end{abstract}

\begin{keyword}
 Opto-hydrodynamics -- Optical radiation pressure --
Boundary integral element method -- Interfacial flow.
\end{keyword}
\end{frontmatter}
\newpage
\section{Introduction}
\label{intro} The deformation of liquid-liquid interfaces by the
optical radiation pressure has received increasing attention in
the past few years as many practical applications of laser-induced
surface deformation are now under development. Among others, we
can cite interfacial characteristics measurements such as viscosity
\citep{yoshitake05} or surface tension \citep{mitani02}, as well
as fluid membranes manipulation with optical tweezers
\citep{guo98,lin01}. Effects of the radiation pressure have been
also recognized as an appealing non-intrusive tool for local
manipulation of liquid or soft materials
 giving birth to many applications in biotechnologies \citep{guck00,moses98}.

Historically, the deformation of fluid-fluid interfaces resulting
from the radiation pressure induced by an impinging focused laser
beam was first identified in the early 70's by Ashkin and Dziedzic
\cite{ashkin73}. Later, Zhang $\&$ Chang \cite{zhang88}
illuminated a $50~\mu$m radius water drop with a $100-200$~mJ
laser pulse and showed strong distortion of the droplet
 surface at its front and rear regions. At low pulse energy, the droplet interface exhibited
oscillations that damped out on time scales of tens of
microseconds, whereas at high energy a disruption generating a jet
of micro-droplets was observed at the rear part of the drop. The
drop distortion was theoretically studied by Lai et al.
\cite{lai89} and later by Brevik et al. \cite{brevik99}. Based on
a linear wave theory adapted to low energy pulses, their analysis
predicted drop oscillations very similar to those observed in the
experiments. However, drop deformations under higher energy
pulses were not modelled as their linear model can no longer be
used to describe the droplet shapes in the regime close to
disruption.

Thus, current existing theoretical descriptions of optically
induced flow and surface deformations are restricted to small
amplitude deformations. However, recent experiments on very soft
interfaces have evidenced several regimes, ranging from the
so-called classical {\it linear regime} for small beam intensity,
in which the height of the deformation linearly depends on the
radiation to Laplace pressure ratio referred to in the following 
as $\xi$ \citep{casner01a}, to {\it nonlinear regimes}
\citep{casner03a,wunenburger06a} with a possible interface breakup
for even larger beam intensities \citep{casner03b}.\ In these
recent experiments, a continuous Ar$^+$ laser wave (wavelength in
vacuum $\lambda_0=514.5$~nm) of waist $\omega_0 \approx 3$ to
$15~\mu$m was used to bend the interface between two liquid phases
in coexistence close to their liquid-liquid critical point. Two
reasons motivated this choice. As the separated phases of these
near-critical mixtures have very low surface tension ($\gamma\sim
10^{-7}$ ~N/m), a laser beam of moderate power $P\sim 1$~W becomes
sufficient to induce large interface deformations of typical size
$\sim 10-100$~$\mu$m. Moreover, near-criticality raises
universality concepts demonstrating the generality of the
purpose. As predicted from the photon momentum balance at the
interface, in the linear regime the radiation pressure induces the
same deformations for upward or downward laser propagation
\citep{casner01a,wunenburger06a}. However, in the nonlinear regime
the invariance of the interface deformation with respect to the
direction of propagation of the wave breaks down. Stable
nipple-like deformations were observed in the case of a
propagation from the less refractive fluid (marked as (1)) to the
more refractive one (marked as (2)) (\textit{upward propagation})
\citep{casner03a}, while in the other case where the laser wave
propagates from the more refractive fluid to the less refractive one (\textit{downward
propagation}), the interface becomes unstable, leading to the
formation of a needle-like deformation emitting micro-droplets
\citep{casner03b} (see Fig. 1).

\begin{figure}[h!!!]
\begin{center}
   \includegraphics[scale=0.65]{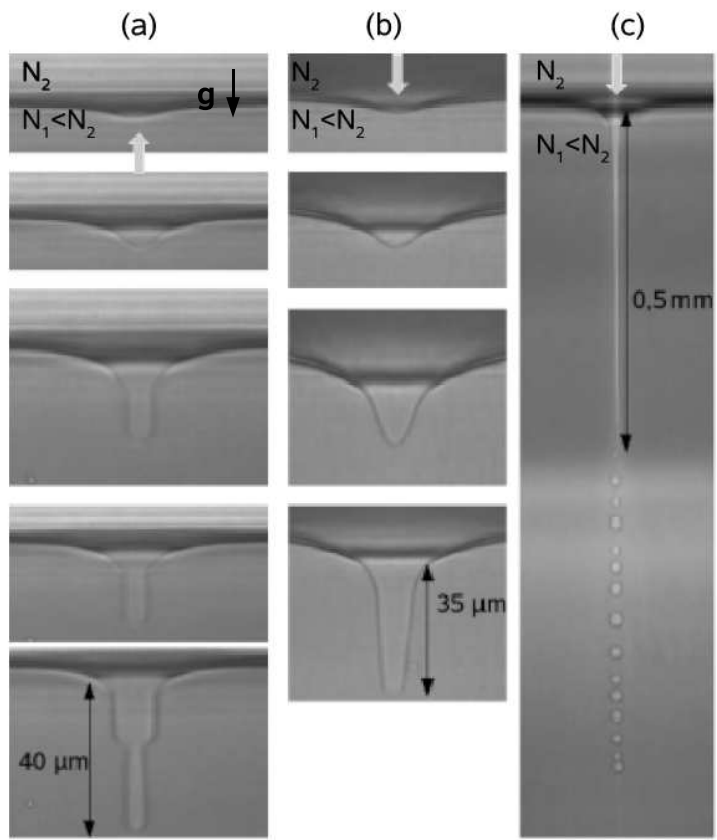}
\caption{Interface deformations induced at $(T-T_C)=3.5K$ by a
laser beam of waist $\omega_0=5.3 \mu m$. (a) Laser propagating
upward from the less refractive fluid to the more refractive fluid
as indicated by the white arrow. P increases from top to bottom
and is successively equal to 120, 240, 360, 390 and 720 $mW$. (b)
Downward direction of propagation. P= 124, 248 and 372 $mW$ from
top to bottom; the bottommost picture (405 $mW$) shows the
destabilization of the interface leading to the formation of a
stationary jet similar to that illustrated in (c).}
\end{center}
\end{figure}

The aim of the present work is to investigate whether the
differences in interface deformation with respect to the direction
of propagation of the wave (called hereafter {\it invariance
breakdown}) is numerically predictable and if the predicted
interface shapes and heights agree with experiments. Our paper is
structured as follows. In Section 2, we briefly describe the
experimental setup, sample properties, and the physical model
for laser-induced interface deformations. This model assumes an
axisymmetric Stokes flow in each liquid and a boundary condition
at the interface describing the competition between viscosity,
optical radiation pressure, capillarity and gravity effects.  A
brief description of the Boundary Integral Element Method (BIEM)
is presented in Section 3, emphasizing the advantages of its
application to interfacial flows. In Section 4, comparisons
between numerical and experimental results are shown and discussed
for both the linear/nonlinear regimes of deformation and both
directions of propagation, illustrating by the way the efficiency
of our approach to investigate the subtle coupling between the
effects of light and flow.

\section{Experimental configuration and physical model}
Exhaustive experimental details on the configuration and protocol
used here were reported earlier \citep{casner01a,wunenburger06a}.
\subsection{Experiments}
In figure 2 we have represented a picture of a typical
liquid-liquid interface deformation induced by a laser beam
propagating upward, together with the notations used throughout
this work. Cylindrical coordinates (${\bf e_r, e_z, e_\alpha})$
with their origin $O$ located at the intersection of the beam axis
with the initial flat interface are chosen for this study and are
shown in Fig. 2. A point $\textbf{x}$ is thus marked by the space
coordinates $(r,z,\alpha)$.
\begin{figure}[h!!!]
\begin{center}
 \includegraphics[scale=0.7]{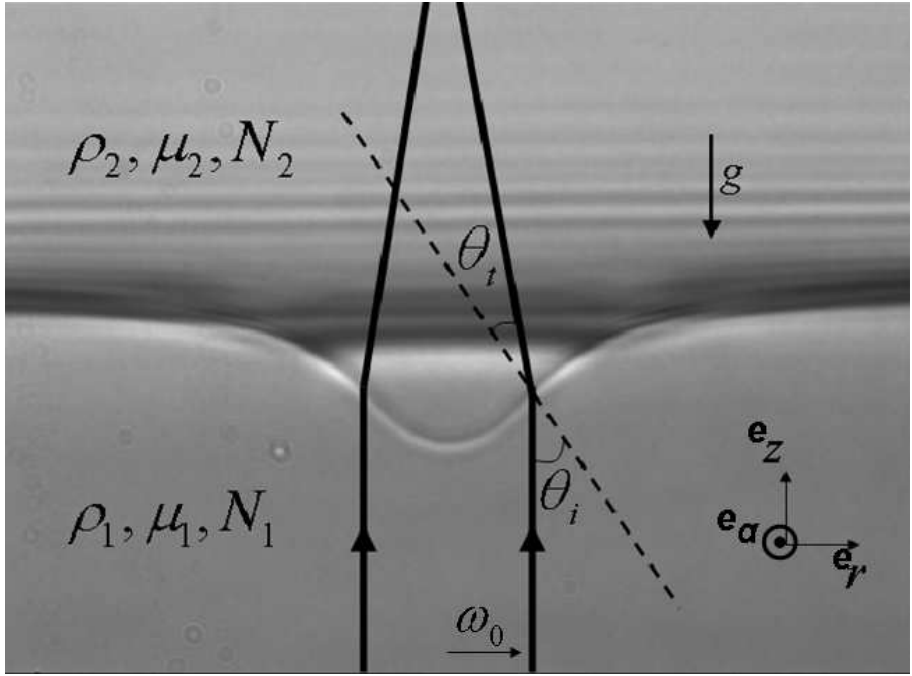}
\caption{Liquid-liquid interface deformation by a focused laser
wave propagating upward. $T-Tc=3.5K$; $\omega_0=5.3 \mu m $ ;
$P=240 mW$. The laser light scattered towards the camera has been
filtered. See text for notations.}
\end{center}
\end{figure}

Incidence and transmission angles of light are respectively
denoted by $\theta_i$ and $\theta_t$.

The two-phase liquid sample is enclosed in a fused-quartz cell of
optical path length $l=2$~mm. The bending of the liquid-liquid
meniscus is driven by a linearly polarized continuous $Ar^+$ laser
(wavelength in vacuum $\lambda_0=514.5$~nm) in the TEM$_{00}$
gaussian mode. The beam is weakly focused on the interface by a
$10\times$ microscope objective to ensure a cylindrical
distribution of the intensity near the meniscus. Thus, in the
vicinity of the liquid-liquid interface, the laser beam intensity $I$ has
the following distribution:
\begin{equation}
I(r,z) \approx I(r)=\frac{2P}{\pi \omega_0^2}
\exp\left(-2\left(\frac{r}{\omega_0}\right)^2\right),
\end{equation}
where $P$ is the laser beam power. The beam waist $\omega_0$ can
be adjusted from 3 to $15~\mu$m. Depending on the setup
configuration, the laser beam propagates vertically either upward
from fluid $(1)$ to fluid $(2)$ or downward from fluid $(2)$ to
fluid $(1)$.

\subsection{Two-phase sample properties}
The investigated fluid-fluid interface is obtained according to the
following procedure. Using a quaternary liquid mixture made of
toluene, sodium dodecyl sulfate, n-butanol and water, we prepare a
water-in-oil micellar phase of microemulsion whose composition is
such that at room temperature it is close to a critical consolute
line. Close to the liquid-liquid critical temperature,
$T_C=308$~K, the critical thermodynamic behavior of the mixture
belongs to the universality class (d=3 n=1) of the Ising model
\citep{freysz94}. For a temperature $T>T_C$ it separates in two
micellar phases of different concentrations $\varphi_1$ and
$\varphi_2$. The use of a near-critical two-phase fluid is
motivated by the fact that significant interface deformations by
electromagnetic radiation, without nonlinear propagation effects
\citep{freysz94} or disturbing thermal coupling, require weak
interfacial tension and buoyancy. With our fluids, both effects
are fulfilled since they vanish when the critical point is neared.
Moreover, the interfacial tension $\gamma$ of phase-separated
near-critical supra molecular fluids is intrinsically smaller than
that of near-critical pure fluids, thus enhancing even more
interface deformations.

The mixture is thermally controlled at a temperature $T$ above
$T_C$. Since the density (index of refraction) of water is larger
(smaller) than that of toluene, the micellar phase of largest
concentration $\varphi_1$, density $\rho_1$ and smallest
refractive index $N_1$ is located below that of lower micellar
concentration $\varphi_2$, density $\rho_2$ and of refractive
index $N_2$. Given the temperature range investigated in the
reported experiments $T-T_C=2$~to~$15$~K, thermophysical
properties of the two-phase sample can be satisfactorily evaluated
using usual asymptotic scaling laws of near-critical
phenomena. For interfacial tension:
\begin{equation}
\gamma=\gamma_0\left(\frac{T-T_c}{T_c}\right)^{2
\nu},\label{gamma}
\end{equation}
with $\gamma_0=10^{-4}$~N/m and $\nu=0.63$. Assuming that
(i) a scaling law accurately describes the variations of $\Delta
\varphi=\varphi_1-\varphi_2$ in the investigated broad temperature domain $T-T_C$ and that (ii) the coexistence curve is symmetric versus the critical concentration $\varphi_c$, the micellar concentration in each
phase can be estimated by:
\begin{eqnarray}
\varphi_1&=&\varphi_c+\frac{\Delta \varphi}{2}, \\
\varphi_2&=&\varphi_c-\frac{\Delta \varphi}{2},
\end{eqnarray}
with $\varphi_c=0.11$ and:
\begin{equation}
\Delta \varphi=\Delta \varphi_c \left( \frac{T-T_c}{T_c}
\right)^{\beta},
\end{equation}
with $\beta=0.325$.

The value of the critical amplitude $\Delta \varphi_c$, can be
estimated theoretically for this system with the main assumption
that our mixture is binary $\Delta \varphi_c=\sqrt{16\pi \varphi_c
R^+}=1.458$ \cite{casner02,langer80}. $R^+=0.37$ being a universal
ratio in microemulsions \cite{mon88}. As micellar phases are in
fact quaternary components fluids, their phase diagram presents
some asymmetry in $\varphi$ leading either to over or
underestimate the critical amplitude $\Delta \varphi_c$ depending
on which side of the coexistence curve is chosen. The theoretical
value of $\Delta \varphi_c$ leads to an overestimation of the
optical index contrast $\Delta N$. Consequently a modified value
for $\Delta \varphi_c$, $0.42$, has been adopted in this study,
based on a quantitative comparison between numerical and
experimental interface steady hump heights in the linear regime of
deformations. This modified value gives acceptable predictions for
the concentrations $\varphi_1$ and $\varphi_2$. In fact, $\Delta
\varphi_c$ should be understood as a free parameter in the
estimation of phases properties.

The density of each phase $\rho_i$, $i=1,2$ can be written as a
function of $\varphi_i$:
\begin{equation}
\rho_i=\rho_{mic}\varphi_i+\rho_{cont}(1-\varphi_i),
\label{densite composite}
\end{equation}
where $\rho_{mic}=1045$~kg.m$^{-3}$ and
$\rho_{cont}=850$~kg.m$^{-3}$ are the densities of the micelles
and continuous phases respectively. At $T-T_C=3.5K$ the estimated
density contrast is $\Delta \rho=\rho_1-\rho_2=\simeq20kg/m^3$.

As the average distance between two micelles is small compared to
the wavelength of the laser wave, the mixture can be regarded as
homogeneous from the electromagnetic point of view. Thus, the
mean-field model for the relative dielectric permittivity
$\epsilon_i$ of the mixture predicts \citep{landau46}:
\begin{equation}
\epsilon_i(\varphi_i)=\varphi_i\epsilon_{mic}+(1-\varphi_i)\epsilon_{cont}-\frac{\varphi_i(1-\varphi_i)(\epsilon_{mic}-\epsilon_{cont})^2}{3(\varphi_i\epsilon_{mic}+(1-\varphi_i)\epsilon_{cont})}.
\label{permittivite composite}
\end{equation}
This relation is used to estimate $N_i$ taking into account the fact that:
\begin{equation}
\epsilon_i=N_i^2~,~ i=1,2.
\label{indices}
\end{equation}
along with $\epsilon_{mic}=1.86$ and $\epsilon_{cont}=2.14$,
$\epsilon_{mic}$ and $\epsilon_{cont}$ being the relative
permittivity of the micelles and continuous phases respectively.

At $T-T_C=3.5K$, the optical indices are $N_1=1.447$ and
$N_2=1.457$.

In addition, since concentrations are weak, we use Einstein's
relation to estimate the dynamic viscosity $\mu_i$ of each phase:
\begin{eqnarray}
\mu_1=\mu_0\left(1+2.5\frac{\Delta \varphi}{2}\right)\\
\mu_2=\mu_0\left(1-2.5\frac{\Delta \varphi}{2}\right),
\end{eqnarray}
with $\mu_0=1.269 $~Pa.s.

Nevertheless, these values could be shifted by possible laser
heating of the fluids. In order to estimate the resulting change
in the physical properties such as interfacial tension $\gamma$ or
viscosity $\mu_i$ due to temperature increase, we consider
the steady diffusion equation in cylindrical coordinate with the
absorbed laser intensity as a source term:
\begin{equation}
\nabla^2 T_I(r)+\frac{\alpha_{th}}{\Lambda_{th}}I(r)=0, \label{diffusion}
\end{equation}
We assume stationary conditions since the thermal diffusion time
scale is much smaller than the viscous one. $T_I(r)$ is the local
increase of temperature due to local heating of the laser wave.
$\alpha_{th}\simeq3~10^{-4} cm^{-1}$ is the thermal absorption and
$\Lambda_{th}=1.28~10^{-3} Wcm^{-1}K^{-1}$ is the thermal
conductivity.
%In order to solve equation (\ref{diffusion}), we consider the boundary conditions $T_I(\omega_I)=0$ with $\omega_I=\sqrt{5\omega_0^2}$ being the length scale where thermal effects are observable \citep{moore65,casner02}. The solution to the equation (\ref{diffusion}) can be achieved using a Fourier-Bessel transform. The local increase in temperature is then given by the following relation: \\
%\begin{equation}
%T_I(r)=\frac{\alpha_{th} P}{4 \pi \Lambda_{th}}[E_1(\frac{2\omega_I^2}{\omega_0^2})-E_1(\frac{2r^2}{\omega_0^2})-ln(\frac{r^2}{\omega_I^2})],
%\end{equation}
%$E_1(x)$ is the one-argument exponential function $[E_1(x)=\int_x^\infty \frac{e^{-k}}{k} dk]$.\\
Using a Fourier-Bessel transform to solve equation (\ref{diffusion}) \citep{moore65,casner02}, we find that the maximum increase in temperature is:\\
\begin{equation}
T_I(r=0) \simeq \frac{\alpha_{th} P}{4 \pi \Lambda_{th}}ln(100\Gamma)
\end{equation}
where $\Gamma=1.781$ is the Euler constant. Considering now
equation (\ref{gamma}), we can estimate the change of interfacial
tension due to the increase in temperature:
\begin{equation}
\frac{\partial \gamma}{\partial T}=\gamma \frac{1.26}{T-T_C}.
\end{equation}
At $T-T_C=3.5K$ and for $P=1W$, we find that
$\displaystyle{\frac{\partial \gamma}{\partial T}\simeq1.3~10^{-7}
Nm^{-1}K^{-1}}$ and $T_I(r=O)\simeq0.1K$ which leads to
$\displaystyle{\frac{\Delta \gamma}{\gamma}\simeq3.6\%}$.
Thermocapillary effects can thus be confidently discarded.

Considering now the viscous dependance on temperature, we use the
following empirical law given for microemulsions \cite{freysz90}:
\begin{equation}
\mu(T)=[1.934-0.019(T-273)]10^{-3}.
\end{equation}
Consequently, $\displaystyle{\frac{\partial \mu}{\partial
T}=-0.019~10^{-3}}$ and thus for $P=1W$ we find
$\displaystyle{\frac{\Delta \mu}{\mu}\simeq2\%}$. This second
estimation ensures negligible thermal effects.

\subsection{Electromagnetic force and pressure}
As expressed in \cite{landau46}, the total electromagnetic force
per unit volume exerted by the laser in each phase is given by:
\begin{equation}
{\bf f_{em}}_i =-\frac{1}{2}\epsilon_0{\bf E_i}^2\nabla\epsilon_i +  \frac{1}{2}\epsilon_0\nabla \left[{\bf E_i}^2\rho_i\frac{\partial\epsilon_i}{\partial\rho_i}\right] + \frac{\epsilon_i-1}{c^2}\frac{\partial}{\partial t}({\bf E_i} \times {\bf H_i}) . \\
\label{fem}
\end{equation}
In this expression, \textbf{E} (respectively \textbf{H}) is the
electric (respectively magnetic) field associated to the laser
wave, $\epsilon_0$ is the permittivity of vacuum and $c=3~10^8m/s$
is the celerity of light in vaccum.\\ The first term, whose jump
across the liquid interface results in the usually called optical
radiation pressure, is due to the change in photon momentum from
one fluid to the other. This momentum change is due to the
discontinuity of permittivities across the interface. As both
fluids are assumed to be homogeneous, the optical radiation term
cancels within each phase and only acts on the interface.

The second term, referred to as the electrostrictive force, also
undergoes a jump at the interface due to the difference in both
the optical properties and the electric fields \textbf{E} between
the two liquids in contact. However, this force also acts as a
bulk force within each phase because of the radial dependence of
the electric field \textbf{E}. We will demonstrate, in adequation
with previous theoretical investigations \citep{lai89,brevik99},
that the electrostriction does not contribute to the motion and
shape of the interface as its bulk contribution is compensated by
its surface one. The quantity
$\frac{\partial\epsilon_i}{\partial\rho_i}$, which depends on the
density of each phase through a nonlinear relationship can be
deduced from equations (\ref{densite composite}) and
(\ref{permittivite composite}).

Finally, in equation (\ref{fem}), the third term is called the
Abraham term, and is undetectable at optical frequencies
\citep{lai89,brevik99} and thus cancels out for our purpose.

In the following, we denote $E^2=<E^2>$ the quadratic value of the
electric field averaged over an optical period.

Assuming incompressible fluids ($\rho_i$ is homogeneous throughout
each phase $i$, $i=1,2$), we can include contributions of the
bulk forces (gravity and electrostriction) in the pressure field
and define a pseudo-pressure $p_i$ given by:
\begin{equation}
 p_i=p_{i0} + \rho_i g z -\frac{1}{2}\epsilon_0 \left({\it E_i}^2
 \rho_i\frac{\partial\epsilon_i}{\partial\rho_i}\right), i=1,2, \label{pression}
\end{equation}
where $p_{i0}$, $i=1,2$, is the pressure in each phase and
$g=9.81$~m.s$^{-1}$ is the acceleration of gravity.

The electromagnetic stress tensor defined by
Landau \citep{landau46} such that $\nabla.{\bf T^{em}_i=f_{emi}}$
can be expressed as:
\begin{equation}
 {\bf T^{em}_i}=\frac{1}{2}\epsilon_0 \left({\it E_i}^2
 \rho_i\frac{\partial\epsilon_i}{\partial\rho_i}\right){\bf I}-\frac{1}{2}\epsilon_0\epsilon_i{\it E_i}^2{\bf I}+\epsilon_0\epsilon_i{\bf E_iE_i}
 , i=1,2. \label{landau}
\end{equation}
In the case of a steady interface, the jump from fluid 1 to
fluid 2 of the local pressure and of the electromagnetic stress
are balanced by the Laplace pressure:
\begin{equation}
(p_{20}-p_{10}){\bf n} + [{\bf T^{em}_1}-{\bf T^{em}_2}].{\bf n}=\gamma\kappa(r){\bf n},
  \label{saut}
\end{equation}
where {\bf n} is the unit vector directed from liquid 1 to liquid
2 and normal to the interface and
$\displaystyle{\kappa(r)=\frac{1}{r}
\frac{d}{dr}\frac{rz'}{\sqrt{1+z'^2}}}$ is the curvature of the
interface in cylindrical coordinates, $z'=\frac{dz}{dr}$ is the
local slope of the interface.

Rewriting Eq. (\ref{saut}) in terms of pseudo-pressures leads to:
\begin{eqnarray}
(p_{2}-p_{1})+(\rho_1-\rho_2)gz+\frac{1}{2}\epsilon_0 ({\it E_2}^2
 \rho_2\frac{\partial\epsilon_2}{\partial\rho_2}-{\it E_1}^2
 \rho_1\frac{\partial\epsilon_1}{\partial\rho_1}) && \nonumber \\ + {\bf n}.[{\bf T^{em}_1}-{\bf T^{em}_2}].{\bf n}=\gamma\kappa(r).
  \label{sautbis}
\end{eqnarray}
Therefore, the bulk contribution of the electrostriction is balanced by its
 surface contribution and the interface shape does not depend on electrostriction.\\
Consequently, the equilibrium equation of the interface can be written as:
\begin{equation}
(p_{2}-p_{1})+(\rho_1-\rho_2)gz-\frac{1}{2}\epsilon_0( \epsilon_1{\it E_1}^2
-\epsilon_2{\it E_2}^2)
  +\epsilon_0(\epsilon_1{\bf E_1E_1}.{\bf n}-\epsilon_2{\bf E_2E_2}.{\bf n}).{\bf
  n}=\gamma\kappa(r).
  \label{equi}
\end{equation}
At final equilibrium, only normal stress act on the interface, consequently there is no flow within the phases and this equilibrium in each phase is characterized by:\\
\begin{equation}
 \nabla p_i=0, i=1,2.\\
\end{equation}
In the experiments, the laser wave was linearly polarized so that
the electric field ${\bf E}= E {\bf e_\alpha}$ was perpendicular
to the plane of observation of azimuthal coordinate $\alpha=0$.
Therefore ${\bf E}$ is continuous across the interface in this
plane.

Conversely, in the plane defined by $\alpha=\pi/2$, the electric
field is within the plane of propagation (parallel polarization)
and thus is no longer continuous across the interface. However, as
shown in \citep{casner03a}, the first term of the electromagnetic
force, which is responsible for the optical radiation pressure
acting on the interface, is quasi-independent of beam polarization
as long as refractive indices are sufficiently close to each
others ($\frac{N_1}{N_2}\sim1$). We can then assume as in a
previous investigation \citep{hallanger05} a circular polarization
of the electric field ${\bf E}$. Therefore, using the definition
of the irradiance $I(r)=\frac{\epsilon_0}{2}N_i c {E_i}^2$
in the case of a beam propagating upward, the radiation pressure
can be written as:
\begin{equation}
\Pi^{up}(r,\theta_i,\theta_t)=-\frac{I(r)}{c}\cos\theta_i(2N_1\cos\theta_i-T^{up}(N_1\cos\theta_i+N_2\cos\theta_t)),
  \label{radup}
\end{equation}
 while in the case of a
downward beam propagation it is given by:
\begin{equation}
\Pi^{down}(r,\theta_i,\theta_t)=\frac{I(r)}{c}\cos\theta_i(2N_2\cos\theta_i-T^{down}(N_2\cos\theta_i+N_1\cos\theta_t)).
  \label{raddown}
\end{equation}
Here, $T^{up}$ and $T^{down}$ are the Fresnel transmission
coefficients in energy. In the case of a circular polarization, they are expressed as:
\begin{eqnarray}
\lefteqn{T^{up}=T^{down}=(2 N_1 N_2 \cos\theta_i \cos\theta_t)}
 \nonumber\\
& \left( \frac{1}{ (N_1 \cos\theta_i+N_2 \cos\theta_t) ^2} +
\frac{1}{(N_2 \cos\theta_i+N_1 \cos\theta_t)^2} \right)
\end{eqnarray}
As a final remark, note that the expression of the electromagnetic
pressure in the case of a downward propagation is valid until
achieving the condition of total reflection
$\theta_i$<$\theta_{TR}=\arcsin \left( \frac{N_1}{N_2}\right)$.
What occurs when light is totally reflected by a highly deformed
interface is out of the scope of the present numerical study.

\subsection{Flow equations and boundary conditions}
We consider a cylindrical domain of radius $R$ and height $H$ as
depicted in Figure 3. We choose to treat all the governing
equations in a dimensionless form by using the laser waist
$\omega_0$ as the characteristic length scale and the viscous
relaxation velocity $u^*=\gamma/\mu_2$ as the reference velocity
associated to the characteristic timescale $\tau^*=\mu_2
\omega_0/\gamma $. The reference pressure is taken as
$p^*=\frac{\mu_i u^*} {\omega_0}$~,~$i=1,2$. Since the weak
absorption of light at laser frequency ensures negligible thermal
effects, we consider all liquid properties ($\gamma$, $\rho_i$,
$\mu_i$, $i=1,2$) constant in adequation with our estimation of
the change in interfacial tension and viscosity due to a local
increase in temperature. Moreover, given the value of the Reynolds
number of the flow $\rho_2 u^* \omega_0/\mu_2 \sim 10^{-3}$,
inertia is negligible compared to viscous forces, allowing a
quasi-steady creeping flow assumption. Thus, flow in each liquid
is governed by Stokes and mass conservation equations respectively
given by:
\begin{eqnarray}
-\nabla p_i + \nabla^2 {\bf u_i}&=&0\\
\nabla . {\bf u_i} &=& 0~,~i=1,2.
\end{eqnarray}
${\bf u_i}$ is the dimensionless velocity in fluid i and $p_i$ is
the dimensionless value of the pseudo-pressure defined in
Eq.(\ref{pression}). The hydrodynamic divergence free stress
tensor ${\bf T_i}$ is:
\begin{equation}
{\bf T_i}= -p_{i} {\bf I} + (\nabla {\bf u_i} + ^t\nabla {\bf
u_i}).
\end{equation}

According to Eq. (\ref{pression}), the normal stress jump across
the interface $S_I$, which is compensated by gravity, optical
radiation pressure and capillarity forces can be expressed in the
following form:
\begin{equation}
(\lambda{\bf T_1}.{\bf n}-{\bf T_2}.{\bf n}).{\bf n} =\kappa(r)
-\Pi(r)- \textrm{Bo} \ z. \label{stressjump}
\end{equation}
In this equation, $\lambda = \mu_1/\mu_2$ is the viscosity ratio
and $\textrm{Bo}= (\rho_1-\rho_2) g \omega_0^2 / \gamma $ is an
optical Bond number which represents the ratio of gravity to
capillary forces, $\Pi(r)$ is the dimensionless expression of the
optical radiation pressure resulting from equations (\ref{radup})
and (\ref{raddown}). We denote $\Pi(r)=\Pi(r)^{up}\omega_0/\gamma$
for the upward direction of propagation and
$\Pi(r)=\Pi(r)^{down}\omega_0/\gamma$ for the downward direction.
In order to quantify the effects of the laser wave on the
interface, we define $\xi$ as the electromagnetic to Laplace
pressure ratio on beam axis at normal incidence:
\begin{equation}
\xi=\Pi{(r=0,\theta_i=0,\theta_t=0)}=\frac{4 P}{\pi c \omega_0
\gamma }\frac{N_1(N_2-N_1)}{(N_2+N_1)}.
\end{equation}
A no-slip condition at the interface along with the fact that
fluids are immiscible implies continuity of the velocity ${\bf u}$
on the interface $S_I$:
\begin{equation}
{\bf u}={\bf u_1} = {\bf u_2} \textrm{ for } {\bf x}\in S_I.
%\label[eq1.11]
\end{equation}
Moreover, we assume the classical no-slip boundary condition on
the lateral, upper and lower solid walls $S_{C1}$ and $S_{C2}$
(see figure 3). This leads to:
\begin{equation}
{\bf u_i = 0} ~\textrm{ for } {\bf x}\in S_{Ci}~,i=1,2.
\end{equation}
In order to reproduce the experimental configuration, $R$ must be
large enough compared to $\omega_0$ to meet the hypothesis of an
infinite extent in the horizontal direction. Several tests were
performed with different values of the dimensionless radius
$\beta=\frac{R}{\omega_0}$ of the computational domain, and a
value $\beta=70$ was found to be large enough to satisfy this
constraint. The interface motion is obtained using a Lagrangian
approach. It consists in tracking each fluid particle on the
interface in its Lagrangian motion according to:
\begin{equation}
 \frac{d{\bf x}}{dt}={\bf u}({\bf x}) \textrm{ for } \textbf{x}\in S_I.
 \label{kin}
\end{equation}
That is, the interface is advected along with the flow until
equilibrium is reached for which velocities normal to the
interface are zero, i.e. ${\bf u(x).n=0}$ for $t \rightarrow
t_{\infty}$.

\section{Numerical method}
A brief description of the numerical algorithm is presented in
this section. For more extensive details on the Boundary Integral
Element Method (BIEM) applied to two-phase axisymmetric flow, the
reader may refer to the review by Tanzosh et al. on the solution
of free surface flow problems using this technique
\citep{tanzosh92}. The BIEM reveals to be an excellent tool to
solve interfacial flow problems with high resolution as reported
in the analysis of flow involving electric and magnetic fields
\citep{sherwood87} or buoyancy
\citep{manga94,koch94}.\\

\begin{figure}[h!!!]
\begin{center}
   \includegraphics[scale=0.8]{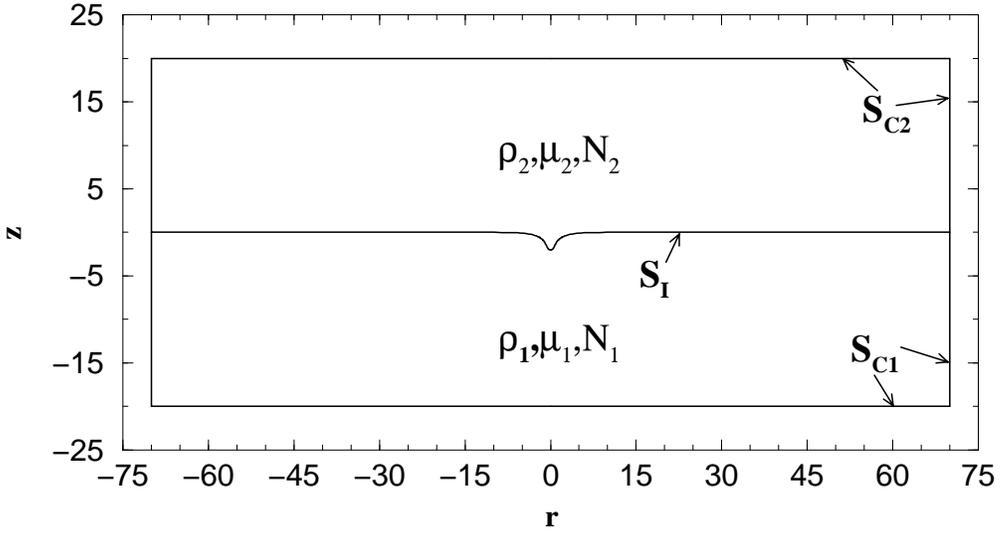}
\caption{Typical configuration of the computational domain showing the two fluids and the liquid interface. The domain is assumed to be axisymmetric along ${\bf e_z}$.}
\end{center}
\end{figure}

Because solutions to the Stokes problem can be formulated in terms
of Green's functions \citep{pozrikidis92}, we can rewrite the
governing equations as a system of integral equations over the
boundaries of the computational domain. Once boundary conditions
on $S_I$, $S_{C1}$ and $S_{C2}$ are used, the two-phase Stokes
problem can be written in the following compact form:
\begin{eqnarray}
\lefteqn{\frac{1+\lambda}{2}{\bf u(x)} = \int_{S_I} {\bf
U.n}(\kappa(r_y)-\Pi(r_y)-Bo z(r_y))dS_y +} \nonumber \\
& & (1-\lambda)\int_{S_I}{\bf n.K.}{\bf
u}dS_y+\lambda\int_{S_{C1}}{\bf U.(T_1.n)}dS_y-\int_{S_{C2}} {\bf
U. (T_2.n)}dS_y. \label{zebigone}
\end{eqnarray}
Here, \textbf{U} and \textbf{K} are Green kernels for velocity and
stress respectively and are given by \citep{pozrikidis92}:
\begin{eqnarray}
{\bf U(d)}&=&\frac{1}{8\pi}(\frac{1}{d}{\bf I}+\frac{{\bf dd}}{d^3}),\\
{\bf K(d)}&=&-\frac{3}{4\pi}(\frac{{\bf ddd}}{d^5}),
\end{eqnarray}
where ${\bf d=x-y}$, ${\bf y}(r_y,z_y)$ is the integration point.
In Eq. (\ref{zebigone}), the first term in the right hand side
describes the flow contribution from interfacial tension,
radiation pressure and gravity, whereas the second term accounts
for shear rates contrast on the interface. This term vanishes when
$\lambda=1$. The third and fourth terms account for shear
occurring on ($S_{C1}$) and ($S_{C2}$) as a result of the no-slip
boundary condition.

Velocities on the interface as well as stress over all the
boundaries $S_{I}$, $S_{C1}$ and $S_{C2}$ are determined by
solving the discrete form of this equation using a numerical
procedure. This procedure requires first the discretization of all
the boundaries $S_{I}$, $S_{C1}$ and $S_{C2}$. Due to integral
formulation and axial symmetry, the problem is reduced to one
dimension and only line boundaries, as represented in Figure 3,
need to be discretized. In this work, the mesh is made of constant
boundary elements i.e. line segments with centered nodes. The
fluid-fluid interface $S_{I}$ is parameterized in terms of arc
length and is approximated by local cubic splines, so that the
curvature can be accurately computed. Distribution and number of
points are adapted to the shape of the interface, so that the
concentration of elements is higher in regions where the variation
of curvature of the interface is large. The number of mesh points
is about 70 for a typical computation of a small interface
deformation. The solid boundaries $S_{C1}$ and $S_{C1}$ are meshed
using about 40 uniformly distributed points. An increase in the
mesh resolution for the interface and the solid boundaries do not
show any significant change in the results.\\Azimuthal integration
of Eq. (\ref{zebigone}) is performed analytically
\citep{lee82,graziani89} reducing Eq. (\ref{zebigone}) to a line
integration which is finally performed using Gauss quadratures
\citep{davis84}. Elliptic integrals resulting from the azimuthal
integration are evaluated using power series expansions
\citep{bakr85}.\\ The motion of the interface is followed using
the kinematic condition (\ref{kin}) which is discretized using an
explicit first-order Euler time scheme. A typical computation
begins with a flat interface at rest. The laser beam is switched
on at $t=0$, and the interface deforms towards fluid $1$ of
smallest refractive index. Computation stops when an equilibrium
state is reached ($\displaystyle{\frac{dz_{(r=0,t)}}{dt}\rightarrow0}$). 
The time step is chosen to be
about 20 times smaller than $\tau^*$.

\section{Results and discussion}
In this section, we compare equilibrium hump heights and interface
shapes obtained experimentally to their numerical predictions in
both linear and nonlinear regimes and for both directions of
propagation. We also present the dynamics of hump formation
expected in the nonlinear regime for both upward and
downward propagations.\\
As mentioned before, few work were dedicated to the modelling of
optical deformations of liquid interfaces in the nonlinear regime.
Hallanger et al. \cite{hallanger05} used a finite difference
method to solve the equation describing the equilibrium state of
the interface for a laser wave propagating from the fluid of
smallest refractive index (here fluid 1) and Wunenburger et al.
\cite{wunenburger06a} used a simple model where gravity was
neglected ($\textrm{Bo}<<1$) to predict the equilibrium hump
heights in the nonlinear regime.\\
\begin{figure}[h!!!]
\begin{center}

 \includegraphics[scale=0.7]{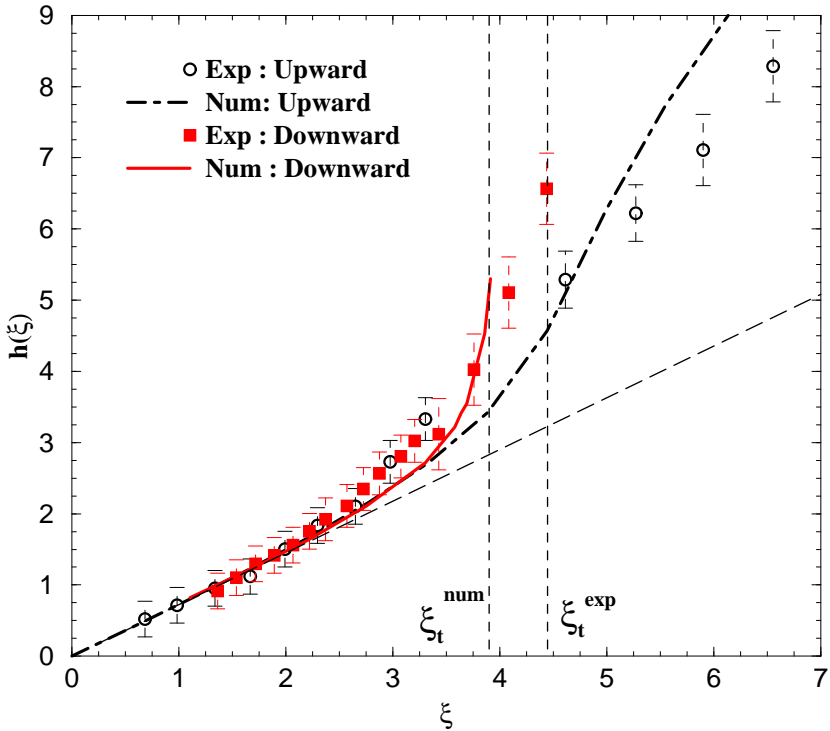}

\caption{Variation of the reduced hump height $h=|z(r=0)|$ versus
the reduced pressure ratio $\xi$ when light propagates upward and
downward for $\omega_0=5.3 \mu m$ and $T-Tc=3.5 K$ ($Bo=0.015$).
Comparison is made between experimental results (symbols) and the
numerical resolution (lines). The bold dashed line is the
prediction for the linear model. $\xi^{exp}_t$ and $\xi^{num}_t$
respectively represent the experimental and numerical thresholds
of instability in the case of downward propagation.}
\end{center}
\end{figure}

In Figure 4, we have represented the variation of the
dimensionless equilibrium interface hump height $h$ as a function
of the pressure ratio $\xi$ for upward and downward propagations.
Experiments (symbols) and numerical predictions (lines) are shown
together.

\subsection{Steady-state deformations, linear regime}
Previous experiments \citep{casner01a,wunenburger06b} performed in
the linear regime of deformations showed that, for $\xi < 2.5$,
$h$ varies linearly with $\xi$ for both directions of propagation
(symbols) and that the linear variation of $h(\xi)$ does not
depend on the direction of propagation for $N_1/N_2 \simeq 1$. A
linear theory assuming $N_1 \simeq N_2$ was proposed
\citep{casner01a,wunenburger06a} which quantitatively agree with
these observations. Indeed, the linear theory gives:
\begin{equation}
z(r)=-\frac{\xi}{4}\int_0^\infty \frac{e^{-k^2/8}}{Bo+k^2}J_0(kr)kdk,\\
\end{equation}
where \textit{$J_0$} is the zeroth-order Bessel \textit{J} function. We deduce:\\
\begin{equation}
h(r=0)=\frac{\xi}{8}e^{\frac{Bo}{8}} E_1(\frac{Bo}{8}),\label{h0} \\
\end{equation}
where $E_1(x)$ is the one-argument exponential function
$[E_1(x)=\int_x^\infty \frac{e^{-k}}{k} dk]$.\\
\textit{In the version published in European Journal of Mechanics, there is an error on the equation (\ref{h0}) where $h(r=0)=\frac{\xi}{2}.... $. This is corrected in this postprint.}

When $N_1$ and $N_2$ are calculated using Eqs. (\ref{permittivite
composite}) and (\ref{indices}), we find that the predicted
$h(\xi)$ variation perfectly agrees with experiments for both
directions of propagation when $\xi < 2.5$ (i.e. in the linear
regime), as shown in Fig. 4 in the particular case $\omega_0=5.3
\mu m$, $T-T_C=3.5K$, i.e. $\textrm{Bo}=0.015$ (dashed straight
line). As shown in Fig. 4, the numerical predictions $h(\xi)$ for
both upward (dot-dashed line) and downward propagations (solid
line) are also found to agree with both the linear theory and
experiments. In addition, Fig. 5 shows good agreement between the
experimental interface shape (dark symbols) and its prediction
using both the linear theory (open square) and the numerical
simulation (solid line) at $\xi=1.35$ for both the upward (Fig.
5a) and the downward propagation (Fig. 5b). This validates the
numerical code versus linear analytical predictions as well as its
accuracy versus experimental results.

\subsection{Steady-state deformations, nonlinear regime}
Experimentally, for $\xi \geq 2.5$, $h(\xi)$ gradually deviates
from linearity. For an upward propagation (open circles in Fig. 4)
$h(\xi)$ increases with a larger slope than that predicted by the
linear theory when $2.5\leq\xi<3.4$. For larger values of $\xi$,
$h(\xi)$ increases again nonlinearly with a slope lying between that
of the linear regime and the above mentioned nonlinear one.

Conversely, in the case of downward propagation (dark
squares), $h$ increases with $\xi$ up to $\xi\simeq3.4$, where its
slope diverges. The divergence of $h'(\xi)$ leads to the jetting
instability studied in details in
\citep{wunenburger06a,casner03b,wunenburger06b} and illustrated in
Fig. 1 .

The invariance breakdown of the hump height below the instability
threshold can be qualitatively explained by considering how the
electromagnetic pressure $\Pi(r,\theta_i,\theta_t)$ varies with
the incidence angle $\theta_i$ for both propagations
\citep{casner03a}. Whereas the electromagnetic pressure, which
varies as $\frac{\cos\theta_i}{\cos\theta_t}$, monotonously
decreases with $\theta_i$ in the case of upward propagation
because $\theta_t=\arcsin(\frac{N_1}{N_2}\sin\theta_i)$,
conversely it continuously increases with $\theta_i$ in the case
of downward propagation because
$\theta_t=\arcsin(\frac{N_2}{N_1}\sin\theta_i)$, as shown in the
inset in Fig. 6. Thus, a downward propagating laser beam exerts an
electromagnetic pressure on a strongly deformed interface (along
which $\sin\theta_i$ can reach values comparable to unity) larger
than an upward propagating beam. The jetting instability occurring
for the downward propagation can be explained by the fact that the
beam propagates in this case from the large to the low refractive
fluid which makes total reflection of light at the interface
achievable \citep{wunenburger06a,casner03b}.

The comparison between numerical simulation results and
experimental data in the nonlinear regime indicates that, in the
case of the upward propagation, a satisfactory agreement is found
regarding $h(\xi)$ (see Fig. 4) for the experimentally
investigated values of $\xi$. Moreover, good agreement regarding
the interface shape is observed up to $\xi \simeq 5$, as shown in
Fig. 5a for the particular value of $\xi=2.9$ and $\xi=4.6$. This
represents a noticeable progress, since previous work on the
subject \citep{hallanger05} qualitatively reproduced interface
shapes (in the case of upward propagation) but did not compare
experiments to numerical predictions.

For $\xi$ larger than 3.5, nipple-like interface shapes, shown in
Fig. 1, are experimentally observed. They are not reproduced by
the numerical simulations suggesting an additionnal coupling or
feedback effect between the exciting beam and the soft interface
that is not investigated in the present work. As also shown in
Fig. 4, in the case of downward propagation, the numerical
simulation qualitatively reproduces the experimentally observed
monotonous behavior of $h(\xi)$ up to the jetting instability
occurring at $\xi=\xi^{num}_t$, as well as the divergence of its
slope at the instability threshold. Instability of the interface
is numerically predicted to occur when the wave undergoes total
reflection and is {\it a priori} focused toward the hump tip, i.e.
when $\theta_i$ reaches $\theta_{TR}$ at the inflection point of
the interface. As a matter of fact, since total reflection and
focusing of the incident wave by the interface is assumed to be
responsible for the jetting instability \citep{casner03a}, we
assume that beyond $\xi^{num}_t$ the interface will actually
become unstable. Moreover, beyond $\xi^{num}_t$ the physical model
used in the numerical model fails, in particular due to Eq.
(\ref{raddown}) because the numerical simulation of the jet
formation and stability  should include the description of the
complex interplay between wave propagation and interface
deformation acting as a wave guide, which is beyond the scope of
the present investigation.\\Still, the discrepancy between the
values of $\xi^{num}_t$ at which total reflection occurs along the
interface ($\xi^{num}_t=3.9$ in Fig. 4) and $\xi^{exp}_t$ just
beyond which the bell-shaped interface actually loses its
stability ($\xi^{exp}_t=4.43$ in Fig. 4) calls for a study of the
interface shape.

\begin{figure}[h!!!]
\begin{center}
 \includegraphics[scale=0.7]{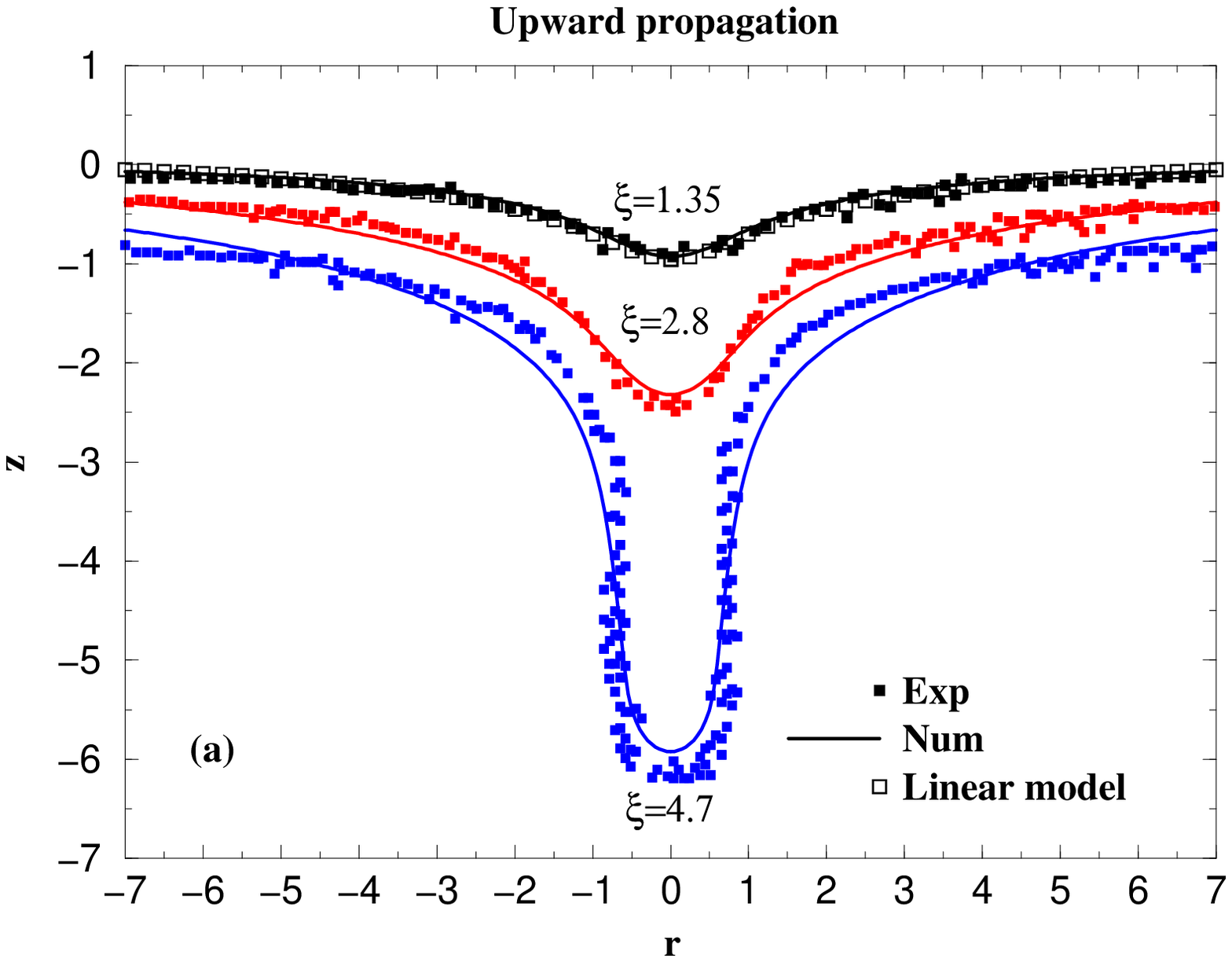}
%\caption{Comparison between experimental and numerical interface profiles for upward (left figure) and downward (right figure) propagation. $\omega_0=5.3 \mu m$ and $T-Tc=3.5 K$ ($Bo=0.015$). }
\end{center}
\end{figure}

\begin{figure}[h!!!]
\begin{center}
 \includegraphics[scale=0.7]{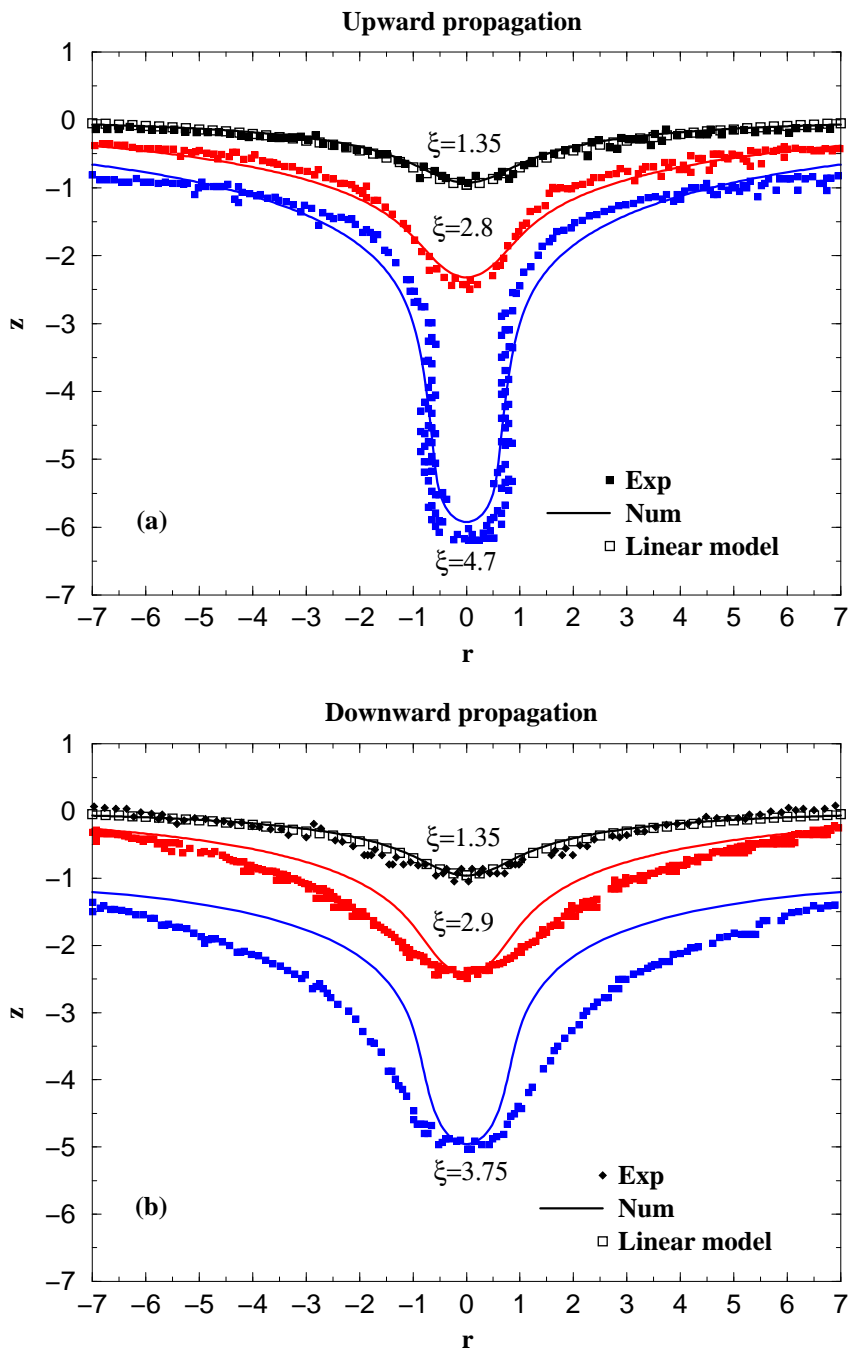}
\caption{Comparison between experimental and numerical interface profiles for upward (a) and downward (b) propagations. $\omega_0=5.3 \mu m$ and $T-Tc=3.5 K$ ($Bo=0.015$). }
\end{center}
\end{figure}

Figure 5b shows that there is also a noticeable discrepancy between
the experimental and numerically predicted interface shapes for
the case of the downward propagation in the nonlinear regime. More
precisely, the experimental hump is wider than the numerically
predicted one, and the experimental slope of the interface shape
is smaller than the numerically predicted one all along the
interface, even when the experimental hump height is well
predicted numerically, as it is the case for $\xi=3.75$ and shown
in Fig. 5b. This observation can explain the discrepancy observed
on $h(\xi)$ obtained experimentally and predicted numerically. As
a matter of fact, since the actual slope of the interface shape is
all along the interface smaller than numerically predicted, the
actual incidence angle $\theta_i$ is also smaller than that
numerically predicted. Consequently, the total reflection of the
wave by the interface reaching a slope equal to $\theta_{TR}$ and
the associated jetting instability are expected to experimentally
occur at a value of $\xi$ larger than that numerically
predicted.

Nevertheless, the reason for this discrepancy between the actual
interface shape and its numerical prediction in the case of a
downward propagation is still unexplained. In Ref.
\citep{wunenburger06a} several possible causes were discussed and
discarded. Among them are the effect of thermocapillary flows due
to light absorption, the role of optical nonlinearities, and the
additional electromagnetic pressure applied on the hump tip due to
the partial reflection of light by the interface below the
instability threshold. A more probable cause may be the viscous
stress applied on the interface by the flow induced by the
scattering of light as a result of refractive index
inhomogeneities occurring in the bulk of each phase
\citep{schroll07}. Indeed, the resulting scattering force is known
(i) to be oriented in the direction of propagation and (ii) of
radial extension much larger than the beam diameter. However, the
exact effect of such a flow on finite size humps, in particular
the interplay between viscous effects on the deformed interface
and the feed back of the hump shape on the flow field in its
vicinity, remains misunderstood and calls for further study.

\subsection{Dynamics of hump formation, nonlinear regime}
Previous work has recently shown that in the linear regime the
interface dynamics is accurately described by a linear theory of
overdamped interfacial circular waves \citep{wunenburger06b}, and
that the interface dynamics is independent of the direction of propagation.\\
However, in the nonlinear regime, the interface dynamics actually
depends on the wave propagation. \\
\begin{figure}[h!!!]
\begin{center}
 \includegraphics[scale=0.7]{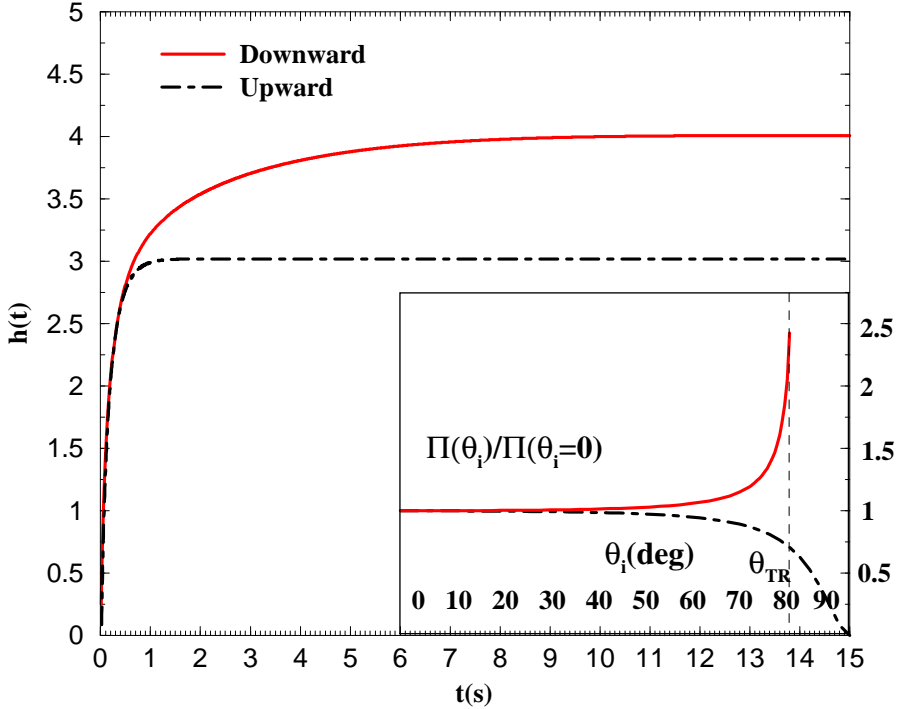}

\caption{Time evolution of the interface hump height for downward
and upward propagation. $\omega_0=5.3 \mu m$ , $T-Tc=3.5K$
($Bo=0.015$) and $\xi=3.75$. Time t is dimensional. The inset
shows the variation of the radiation pressure normalized by its
value at normal incidence, versus the incidence angle for both
propagations. The dashed line (inset) shows the total reflection
threshold $\theta_{TR}$.}
\end{center}
\end{figure}

In Fig. 6, the time evolution of the hump height $h(t)$ is plotted
for both the upward (dot-dashed line) and downward propagations
(solid line) for $\xi=3.75$, a value belonging to the nonlinear
regime. We note that (i) both curves coincide during the first
$500$~ms of the dynamics and eventually separates, (ii) the
transient is significantly shorter in the case of upward
propagation and the equilibrium state is reached much faster.
 The first observation can be explained by noting that
for $t<500$~ms both interface deformations are small enough to be
described by a linear theory which is precisely independent of the
beam propagation. The second observation can be qualitatively
explained by considering the increase of the electromagnetic
pressure with the incidence angle $\theta_i$ in the case of
downward propagation (as shown in the inset of Fig. 6). As
$\theta_i$ increases along the interface the associated
overpressure applying on the growing hump increases its final
height and thus increases the delay for the hump to reach equilibrium. 
On the contrary, the fact that the
electromagnetic pressure decreases when $\theta_i$ increases in
the case of an upward propagation leads to the opposite
conclusion. Consequently, the analysis of the dynamics shows that,
beyond the regime of linear deformations, our numerical approach
becomes predictive for the complex nonlinear regime, thus calling
for new experimental investigation and providing indications on
how to perform these experiments.

\section{Conclusions}
A Boundary Integral Element Method has been used to simulate the
deformation of a fluid-fluid interface induced by a focused laser
beam. Since interface deformations with azimuthal invariance were
expected, an axisymmetric model has been developed. Comparisons
between numerical predictions and experimental data showed two
main results. First, our model can satisfactorily reproduce the
variations of interface equilibrium hump height $h$ with respect
to the ratio of electromagnetic to capillary pressure $\xi$ in 
regimes of small as well as of large deformations. In the case of
an upward propagation, nipple-like interface shapes were not
reproduced, while in the case of a downward propagation,
instability threshold was overestimated probably because actual
interface shapes are wider than numerically predicted. This could
be due to bulk steady flows induced by light scattering as a
result of liquid index inhomogeneitiy which deserves further
experimental characterization before numerical implementation. A
study of the dynamics of the interface for both directions of
propagation was also performed showing a shorter characteristic
time for the interface reaching its steady state in the case of an
upward propagation. Future work should take into account the total
reflection and focusing of light by the deformed interface in
order to potentially predict hydrodynamic instability leading
to jet formation and micro-droplet emission at the tip.
Another interesting investigation would be to study the effect of
light scattering induced flow \citep{schroll07} on jetting
instability. Our first results are thus very encouraging since
they illustrate the accuracy of our numerical model to describe
and predict intriguing properties of nonlinear behaviors of the
coupling between light and liquid interfaces, a subject of
increasing interest due to its wide range of applications in soft
matter physics on the one hand and scarce theoretical results on
the other hand.

\textbf{ Acknowledgements}\\

This research was supported by Centre National de la Recherche
Scientifique (France), Universit\'e Bordeaux 1, and Conseil
R\'egional d'Aquitaine.

%\bibliographystyle{unsrt}
%\bibliography{hamza_min}

\end{document}